\documentclass[prb,a4paper,12pt,onecolumn,superscriptaddress,amsmath,amsfonts,amssymb,preprintnumbers,citeautoscript]{revtex4}

\pdfoutput=1
\usepackage[T1]{fontenc}\usepackage[latin1]{inputenc}
\usepackage{graphicx,booktabs,microtype,afterpage,sidecap, times, soul}
\usepackage[charter]{mathdesign}

\usepackage[colorlinks,plainpages=false,linkcolor=black,urlcolor=blue,citecolor=black,pdfpagemode=UseNone,pdfstartview=FitBH]{hyperref}

\usepackage{graphicx}   
\usepackage{verbatim}   
\usepackage{color}      
\usepackage{subfigure}
\usepackage{gensymb}
\usepackage{textcomp}
\usepackage{hyperref}   
\usepackage{longtable}
\usepackage{csquotes}
\usepackage{commath}
\usepackage[english]{babel} 
\usepackage{multirow} 
\usepackage{cleveref} 
\usepackage[normalem]{ulem} 
\usepackage{xcolor} 
\usepackage{bm}
\usepackage{color}
\usepackage{epstopdf}

\makeatletter
\renewcommand{\@evenfoot}{\hfill\raisebox{-1em}{\bf\thepage}\hfill}
\renewcommand{\@oddfoot}{\hfill\raisebox{-1em}{\bf\thepage}\hfill}
\makeatother

\newcommand{\bna}{\ensuremath{\mathrm{BaNi_2As_2}}\;}
\newcommand{\bnap}{\ensuremath{\mathrm{BaNi_2(As_{1-x}P_x)_2}}\;}
\newcommand{\eg}{$\rm{E_{g}}\;$}
\newcommand{\ega}{$\rm{E_{g,1}}\;$}
\newcommand{\egb}{$\rm{E_{g,2}}\;$}
\newcommand{\bg}{$\rm{B_{1g}}\;$}
\newcommand{\bgg}{$\rm{B_{2g}}\;$}
\newcommand{\ag}{$\rm{A_{1g}}\;$}

\newcommand{\icm}{\,$\rm{cm^{-1}}$}
\newcommand\redsout{\bgroup\markoverwith{\textcolor{red}{\rule[0.5ex]{2pt}{0.4pt}}}\ULon}
\begin{document}

\title{\large{\textbf{An Electronic Nematic  Liquid in $\mathbf{BaNi_2As_2}$}}}

\author{Yi Yao}
\affiliation{Institut f\"{u}r Quantenmaterialien und -technologien, Karlsruher Institut f\"{u}r Technologie, 76021 Karlsruhe, Germany\\}

\author{Roland Willa}
\affiliation{Institut f\"{u}r Theorie der Kondensierten Materie, Karlsruher Institut f\"{u}r Technologie, 76131 Karlsruhe, Germany \\}

\author{Tom Lacman}
\affiliation{Institut f\"{u}r Quantenmaterialien und -technologien, Karlsruher Institut f\"{u}r Technologie, 76021 Karlsruhe, Germany\\}

\author{Sofia-Michaela Souliou}
\affiliation{Institut f\"{u}r Quantenmaterialien und -technologien, Karlsruher Institut f\"{u}r Technologie, 76021 Karlsruhe, Germany\\}

\author{Mehdi Frachet}
\affiliation{Institut f\"{u}r Quantenmaterialien und -technologien, Karlsruher Institut f\"{u}r Technologie, 76021 Karlsruhe, Germany\\}

\author{Kristin Willa}
\affiliation{Institut f\"{u}r Quantenmaterialien und -technologien, Karlsruher Institut f\"{u}r Technologie, 76021 Karlsruhe, Germany\\}

\author{Michael Merz}
\affiliation{Institut f\"{u}r Quantenmaterialien und -technologien, Karlsruher Institut f\"{u}r Technologie, 76021 Karlsruhe, Germany\\}

\author{Frank Weber}
\affiliation{Institut f\"{u}r Quantenmaterialien und -technologien, Karlsruher Institut f\"{u}r Technologie, 76021 Karlsruhe, Germany\\}

\author{Christoph Meingast}
\affiliation{Institut f\"{u}r Quantenmaterialien und -technologien, Karlsruher Institut f\"{u}r Technologie, 76021 Karlsruhe, Germany\\}

\author{Rolf Heid}
\affiliation{Institut f\"{u}r Quantenmaterialien und -technologien, Karlsruher Institut f\"{u}r Technologie, 76021 Karlsruhe, Germany\\}

\author{Amir-Abbas Haghighirad}
\affiliation{Institut f\"{u}r Quantenmaterialien und -technologien, Karlsruher Institut f\"{u}r Technologie, 76021 Karlsruhe, Germany\\}

\author{J\"org Schmalian}
\affiliation{Institut f\"{u}r Quantenmaterialien und -technologien, Karlsruher Institut f\"{u}r Technologie, 76021 Karlsruhe, Germany\\}
\affiliation{Institut f\"{u}r Theorie der Kondensierten Materie, Karlsruher Institut f\"{u}r Technologie, 76131 Karlsruhe, Germany \\}

\author{Matthieu Le Tacon*}
\affiliation{Institut f\"{u}r Quantenmaterialien und -technologien, Karlsruher Institut f\"{u}r Technologie, 76021 Karlsruhe, Germany\\}
\date{\today}

\begin{abstract}
\center\bigskip\thispagestyle{plain}
\begin{minipage}{\textwidth}\textbf{Understanding the organizing principles of interacting electrons and the emergence of novel electronic phases is a central endeavor of condensed matter physics.
Electronic nematicity, in which the rotational symmetry in the electron fluid is broken while the translational one remains unaffected, is a prominent example of such a phase. It has proven ubiquitous in correlated electron systems, and is of prime importance to understand the physics of the Fe-based superconductors. Here, we find that fluctuations of such broken symmetry are exceptionally strong in non-magnetic \bnap, the charge analog to the Fe-based systems, at temperatures well above that of any phase transition. 
This provides evidence for a new type of electronic nematicity, dynamical in nature, which exhibits an unprecedented  coupling to the underlying crystal lattice. Quantum and thermal fluctuations between degenerate configurations cause a splitting of phonon lines, without lifting degeneracies or breaking the rotational symmetry. This nematic liquid is reminiscent of spin liquids in quantum magnetism, offering the perspective that the absence of ordinary  order may  lead to states with long-range quantum entanglement, fractionalized excitations, or unconventional superconductivity.}
\end{minipage}
\end{abstract}

\maketitle\thispagestyle{empty}\clearpage

The normal state of unconventional superconductors generally exhibits a variety of exotic electronic states emerging out of the interplay between intertwined orders. It is at least as
intriguing as the superconducting state itself.
The electronic nematic state is one such exotic state, which has proven particularly insightful in unveiling the properties of Fe-based superconductors. In these materials, nematicity is a canonical example of 
vestigial order which grows out of the magnetic fluctuations of two degenerate magnetic ground states~\cite{Fernandes_ARCMP2019}. 
It has been proposed that the superconducting pairing is enhanced - even possibly mediated - by quantum critical nematic fluctuations~\cite{Lederer_PRL2015}, but their  most prominent effect is rooted in their coupling to the crystal lattice. By softening the $C_{66}$ shear modulus in \textit{e. g.} Ba(Fe$_{1-x}$Co$_{x}$)$_2$As$_2$, this coupling ultimately yields a  lattice distortion at a structural tetragonal-to-orthorhombic phase transition above the superconducting dome~\cite{Boehmer_CRP2016, Chu_PRB2009} .

This coupling to the lattice changes the phonon spectra and dispersion, which in turn provides new routes to probe electronic nematicity. In the fluctuating regime, it was  recently shown that the spatial dependence of the nematic fluctuations can directly be inferred from the softening of acoustical phonons~\cite{Weber_PRB2018, Merritt_PRL2020, Wu_PRL2021} at small but finite momentum ({\bf q} $\neq$ 0).At the Brillouin zone center ({\bf q} = {\bf 0}), the largest effects are observed in the ordered phase, through the lifting of the degeneracy of the $a$- and $b$-axis polarized in-plane vibrations of the square FeAs lattice with \eg symmetry. The resulting relative splitting $\Delta \omega / \omega$ of the modes in the orthorhombic phase can be as large as $8\%$ in the Fe-based superconductors' parent compounds such as BaFe$_2$As$_2$~\cite{Chauviere_PRB2009, Ren_PRL2016, Baum_PRB2018} or EuFe$_2$As$_2$~\cite{Zhang_PRB2016}, exceeding by far the  expectation based on the small orthorhombicity $\delta = \frac{a-b}{a+b} \sim 10^{-3}$. The much weaker effects reported in non-magnetic FeSe~\cite{Hu_PRB2016} suggests that the coupling of nematic degrees of freedom to the lattice in Fe-based superconductors primarily occurs through the spin channel rather than through the orbital one~\cite{Hu_PRB2016}.

At room temperature, \bna has a tetragonal crystal structure (space group I4/mmm) similar to BaFe$_2$As$_2$, but unlike its Fe-counterpart, is superconducting, albeit below a modest critical temperature $T_c \sim$ 0.6~K~\cite{Ronning_JPCM2008}. While earlier  electronic structure studies concluded  low electronic correlations in this system, pointing at conventional phonon-mediated BCS superconductivity~\cite{Subedi_PRB2008, Kurita_PRL2009},  more recent investigations advocate for an exotic normal state, which exhibits a manifold of charge density waves (CDW) instabilities and structural phase transitions interesting in their own right \cite{Lee_PRL2019, Frachet2022, Meingast2022, Merz_arxiv2020, Eckberg_PRB2018, Eckberg_NatPhys2020, Grigorev_arxiv2021, Lee_arxiv2021}, and possible nematic-driven superconducting pairing~\cite{Lederer_PRR2020}. No long-range magnetic order has been reported  so far, and it has been argued that CDW  plays a role similar to that of magnetism in the Fe-based superconductors, making \bna a 'charge analogue' of BaFe$_2$As$_2$. 

Here we investigate the lattice and electron dynamics of \bnap, and report on an exceptionally large splitting of the doubly degenerate Raman active planar vibrations of the NiAs tetraedra. In sharp contrast to the behavior in the iron-based systems, where this splitting was taken as evidence for nematic symmetry breaking, in \bnap it occurs well above any reported structural phase transition temperatures. This calls for a distinction between the lifting of a degeneracy and a dynamical spectral splitting. We show that our observation can be accounted for by a particularly strong coupling of electronic \bg nematic fluctuations, likely of orbital nature~\cite{Merz_arxiv2020}, to the lattice degrees of freedom in this material. Hence, the tetragonal phase of \bna hosts a nematic quantum liquid phase exhibiting dynamic correlations. As we will show, the dynamic splitting can be described in terms of  an entangled superposition of the two degenerate Ising-nematic states that are coupled to a cloud of vibrational quanta. The rich physics of spin liquids - dynamical states without long-range magnetic order but long-range entanglement -  suggests that similar behavior is to be expected in  nematic liquids.

\begin{figure*}[!htbp] 
\centering
\includegraphics[width=\columnwidth]{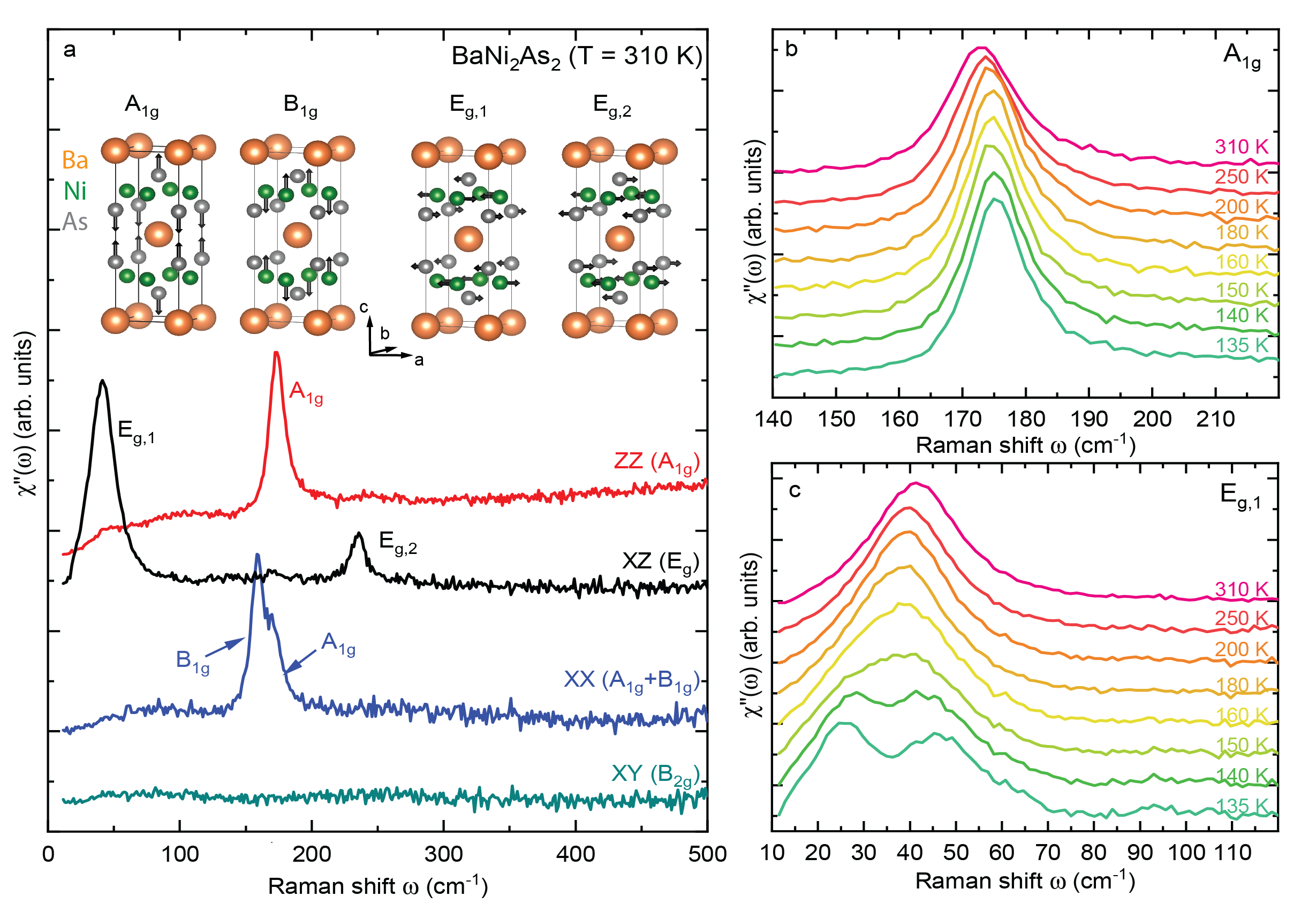}
\caption{\label{fig1} {\textbf{Raman scattering from \bna}} (\textbf{a}) Raman active phonons of \bna, and room temperature Raman spectra obtained in the different incoming and scattered photon polarizations. Detailed view of the temperature dependencies of the \ag  (\textbf{b}) and \ega (\textbf{c}) phonons above $T_{Tri}$ in \bna.}
\end{figure*}

\paragraph*{Experimental results} 
From a point group analysis follows that the tetragonal phase of \bna hosts four Raman-active optical phonons of \ag, \bg and \eg symmetry at the Brillouin zone center. The corresponding eigen-displacements are shown on fig.~\ref{fig1}-a. In this figure, we further report on room temperature Raman scattering measurements performed on BaNi$_2$As$_2$ single crystals. The experiments were carried out in backscattering geometry with $XZ$, $ZZ$, $XX$ and $XY$ configurations, where the first (resp. second) letter refers to the orientation of the incident (resp. scattered) light polarization with respect to the axis of the tetragonal unit cell (Supplementary Information). All four Raman active optical phonon modes were detected. The \ag mode is seen in the $ZZ$ configuration at 172.9 \icm, as well as in the $XX$ channel, where it partially overlaps with the  \bg mode, at 158.6 \icm. The two modes observed in the $XZ$ channel are the doubly degenerate \eg modes referred to as \ega (41.4 \icm) and \egb (235.2 \icm).  With the notable exception of the lowest \ega mode, these energies are in good agreement with the predictions of {\it ab initio} calculations (see Supplementary Information). These calculations also allowed us to estimate the strength of the electron-phonon coupling for the different modes and revealed that the phonon exhibiting the largest coupling is the \ag mode, which consistently displays a weak Fano asymmetry. On the other hand, despite the rather modest calculated EPC, the \ega mode is very broad (full-width-at-half-maximum (FWHM) $\sim$ 22\icm) at room temperature, indicating additional decay channels. 

\begin{figure*}[!htbp]
 \centering
\includegraphics[width=0.7\columnwidth]{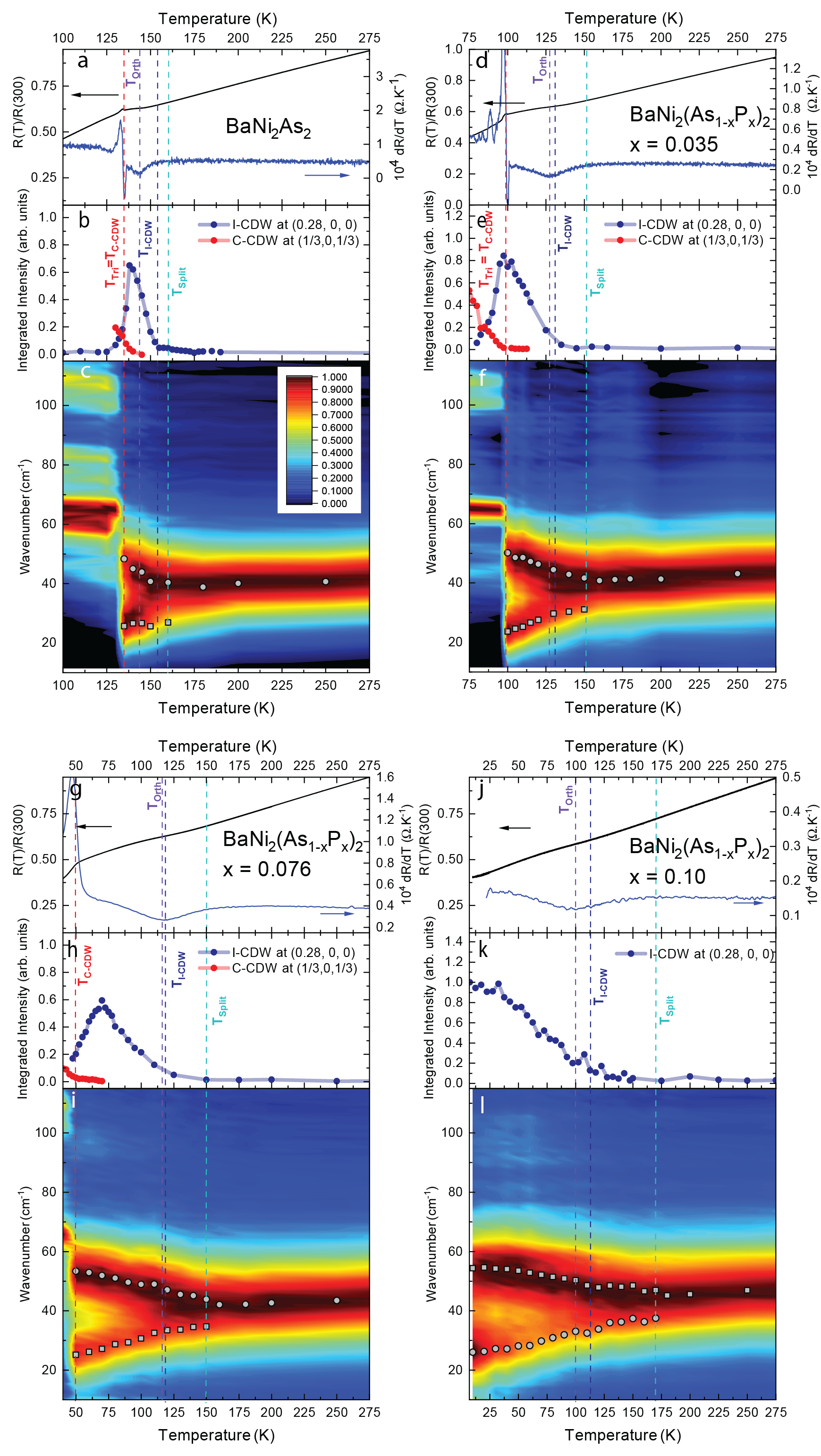}
\caption{\label{fig2} {\textbf{Doping dependence}} (\textbf{a}) temperature dependence of the resistivity and its derivative measured upon cooling in \bna, \textbf{b} of the integrated intensity of the $q_{I-CDW}$=(0.28,0,0) and $q_{C-CDW}$=(1/3,0,1/3) superstructure peaks (measured upon cooling) in \bna (\textbf{c}) temperature dependence of the \ega phonon intensity in \bna (after background subtraction and bose correction, Supplementary Information). (\textbf{d, e, f}) same as (\textbf{a, b, c}) for \bnap (x=3.5\%) (\textbf{g, h, i}) same as (\textbf{a, b, c}) for \bnap (x=7.6\%) (\textbf{j, k, l}) same as (\textbf{a, b, c}) for \bnap (x=10\%)). 
}
\end{figure*}

The singular behavior of the \ega phonon  is confirmed upon cooling. The conventional behavior of the \ag (Fig.~\ref{fig1}-b) and \bg (not shown) phonons is to harden and narrow at low temperatures. In contrast, the \ega mode initially softens upon cooling, starts broadening at 180 K and splits below $T_{Split}$ = 160 K, where two peaks can be resolved; see  Fig.~\ref{fig1}-c.
Just above the first-order transition~\cite{Ronning_JPCM2008} to a triclinic phase at $T_{Tri}$=133 K (on cooling), within which the phonon spectra qualitatively change (Supplementary Information and Fig.~\ref{fig2}), the splitting is as large as 22\icm, that is, more than 50\% of the mode's original frequency.

The degeneracy of an \eg phonon can only be lifted if the four-fold symmetry of the Ni planes is broken. This occurs across the tetragonal-to-orthorhombic structural transition in the Fe-based compounds Ba(Fe$_{1-x}$Co$_{x}$)$_2$As$_2$ ~\cite{Chauviere_PRB2009, Ren_PRL2016, Baum_PRB2018}, EuFe$_2$As$_2$~\cite{Zhang_PRB2016} or FeSe~\cite{Hu_PRB2016}, where \eg modes split into B$_{2g}$ and B$_{3g}$ modes. The largest reported splitting in AFe$_2$As$_2$ (A=Ba or Eu) is $\sim$10\icm ($\sim$8\% of the mode frequency)~\cite{Chauviere_PRB2009, Ren_PRL2016, Baum_PRB2018,Zhang_PRB2016}, significantly larger than in FeSe ($\sim$2.6\icm)~\cite{Hu_PRB2016}. In both cases, this splitting is already considered unusually large, in the sense that it exceeds the expectation based on the lattice distortion. The splitting in our measurements is quantitatively much larger and moreover occurs long before (up to $25 {\rm K}$ ) the four-fold symmetry breaking takes place.

Before discussing the doping dependence of the effect in \bna, we briefly review the potential sources of symmetry breaking that could yield a splitting of the \ega phonon. As FeSe, \bna does not exhibit any magnetic order but a unidirectional, biaxial, incommensurate CDW (I-CDW) above $T_{Tri}$ has recently been reported~\cite{Lee_PRL2019, Eckberg_NatPhys2020, Merz_arxiv2020}. We performed a detailed temperature dependent x-ray diffraction (XRD) investigation of the intensity of the CDW satellite at $q_{I-CDW} = (\pm0.28, 0 ,0)$ (Fig.~\ref{fig2}-a. Note that throughout the manuscript, we will only refer to reciprocal lattice vectors in the tetragonal unit cell). A very weak diffuse scattering signal can be tracked up to room temperature, but a strong increase of the peak intensity is only observed below $T_{I-CDW}$ = 155K. 

It is followed by an orthorhombic distortion of the lattice at $T_{Orth} = 142 $K which can be detected through high-resolution dilatometry~\cite{Merz_arxiv2020} and also manifests itself as a minimum in the derivative against temperature of the resistivity (Supplementary Information), reported in Fig. 2a. In contrast to more pronounced distortions, the identification of the twin structure associated to the structural change was limited in our XRD measurements to a broadening of high order Bragg reflections (e.g. (8,0,0)). This turns to our advantage as it allows us to put an upper bound on the corresponding lattice distortion $\delta = \frac{a-b}{a+b} \sim 10^{-4}$, an order of magnitude weaker than that reported in BaFe$_2$As$_2$. 

Cooling further down, the CDW superstructure peak is suppressed and a commensurate CDW (C-CDW) signal appears at $q_{C-CDW} = (\pm1/3, 0 ,\pm1/3)$ as the system enters the triclinic phase.  We did not detect any additional phase transition in the temperature range at which the \ega mode splitting onsets, and can already conclude at this stage that this splitting is occurring in the tetragonal I4/mmm phase. 

Our first principle calculations confirmed that the amplitude of the orthorhombic structural distortion is in all cases much too small to account for the gigantic energy splitting of the \ega phonons reported here (Supplementary Information). Furthermore, in stark contrast to Fe-based materials~\cite{Chauviere_PRB2009, Hu_PRB2016}, the \ega mode splitting increases linearly and does not show any sign of saturation down to $T_{Tri}$.

We confirmed this by substituting arsenic by phosphorus. This has previously been reported to suppress the triclinic transition~\cite{Kudo_PRL2012}, and to enhance the orthorhombic distortion~\cite{Merz_arxiv2020}. In Fig.~\ref{fig2}, we show the results for \bnap~ with x=3.5\% ($T_{Tri}$= 95 K) and x=7.6\% ($T_{Tri}$= 55K), for which we observe a similar splitting of the \ega mode, which increases linearly as temperature decreases reaching almost 30 \icm at $T_{Tri}$ ($\sim$ 65\% of the mode frequency). Upon further increase of the P-concentration (x=10\%), for which the triclinic (and therefore the C-CDW) transition is completely suppressed, the maximal amplitude of the splitting is reduced and appears to saturate at lowest temperatures. 
In the investigated doping range, the temperature below which the \ega mode splits decreases linearly with doping at a modest rate of $\sim d T_{Split}/d x \sim -2.5K/\%$, comparable to that of $T_{Orth}$ but much lower than $T_{Tri}$, and increases again beyond the suppression of the triclinic phase. The suppression of $T_{Tri}$ further allowed us to resolve a splitting for the \egb mode as well, which amounts to $\sim$ 15\icm at 50 K and could not be resolved at lower P-contents (not shown).

\begin{figure*}[!htbp]
 \centering
\includegraphics[width=\columnwidth]{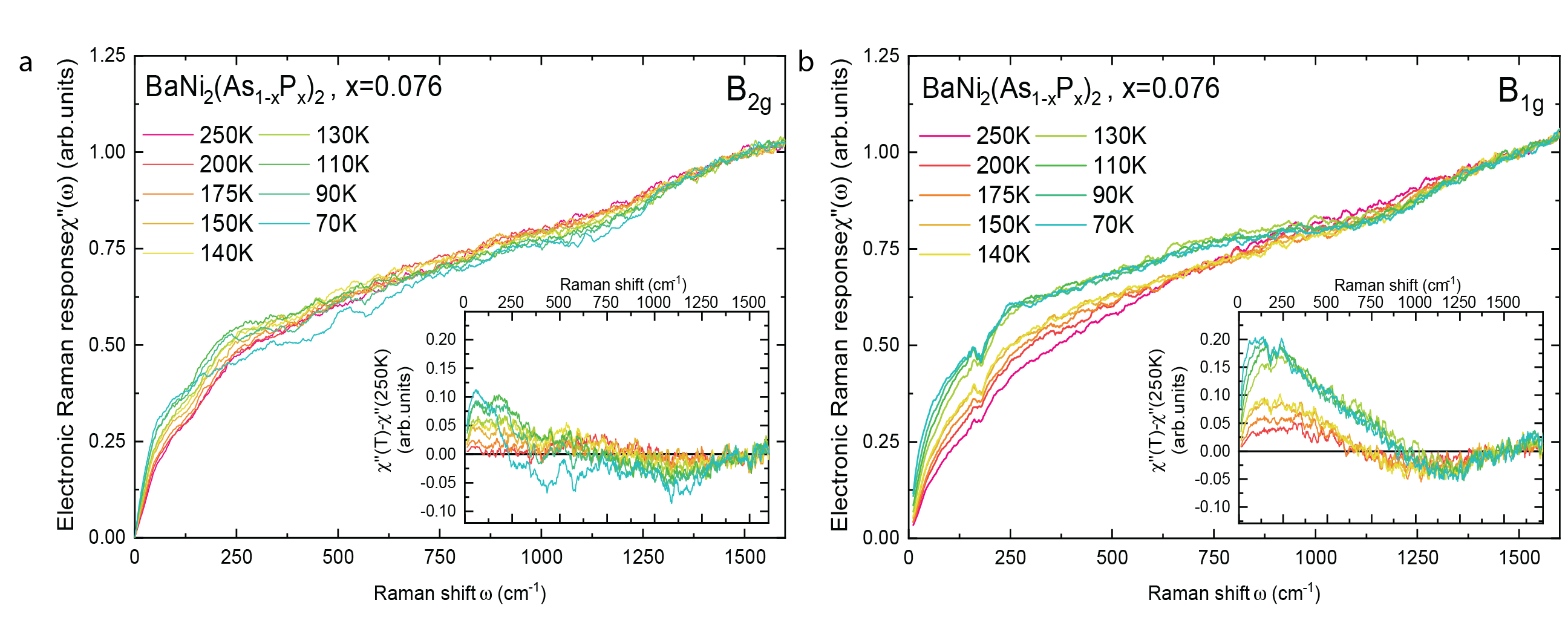}
\caption{\label{fig3} {\textbf{Electronic Raman Scattering}} (\textbf{a}) \bgg electronic response of \bnap (x=7.6\%) \textbf{b}) \bg electronic response of \bnap (x=7.6\%) 
}
\end{figure*}

The observation of the \eg phonon splitting in \bna above any lowering of the symmetry of the compound, strongly suggests a coupling of the mode to fluctuations. Given the lack of magnetism in \bna, orbital degrees of freedom are the most likely candidates. This can directly be tested using electronic Raman scattering, which is a particularly sensitive probe of the charge fluctuations. In Fig.~\ref{fig3} we show the temperature dependence of the electronic Raman response in the \bg and \bgg channels for the $x=6.5\%$ sample (for clarity, the Raman active phonon has been subtracted from the \bg spectrum - Supplementary Information).  In both channels, the electronic response consists of a broad continuum, extending up to 1500\icm, akin to the particle-hole excitations seen in many correlated metals~\cite{Devereaux_RMP2007, Sen_NatCom2020, Kretzschmar_NatPhys2016}. It apparently displays a conventional metallic behavior, with a smooth increase of the low frequency ($\leq$ 250 \icm) response upon cooling, reflecting the decrease of the quasiparticle scattering rate $\Gamma$ (which is inversely proportional to the slope of the Raman response, $\chi^{\prime\prime}(\omega)/\omega|_{\omega \rightarrow 0}$ in the static limit). The main difference between the two channels is quantitative: the low energy \bg intensity gain spans over a broader energy range ( $\geq$ 500 \icm) than the \bgg, it is overall larger and accelerates significantly below $T_{Split}$ $\sim$ 140 K. This observation is in line with recent elasto-resistivity measurements~\cite{Eckberg_NatPhys2020, Frachet2022} and can be interpreted as a signature of \bg nematic fluctuations in \bna. In contrast to the electronic nematicity of the Fe-based superconductors, observed in the \bgg channel, the respective response in \bna appears overdamped, suggesting a strong coupling of the lattice to the nematic fluctuations. 

Having established the presence of nematic fluctuations coupled to the lattice, we can understand the 'splitting' of the \eg phonon spectrum, in the absence of a phase transition that would lift the degeneracy, as a dynamic phenomenon.
To this end, we introduce a minimal model, in which a doubly degenerate lattice vibration couples to an Ising variable $|s\rangle \in \{ |\!\Uparrow \rangle = (1, 0), |\!\Downarrow \rangle = (0, 1) \}$ that describes an electronic nematic degree of freedom, such as an orbital polarization. The two degenerate \eg phonons are described by 
\begin{align}
   H_{E_g} &= \frac{p^2_x}{2M} + \frac{p^2_y}{2M} + \frac{M \omega^2_0}{2}(u^2_x + u^2_y)\, ,
\label{eq: Heg}
\end{align}
where $p_j$ ($j = x,y$) are the phonon momentum operators, while $u_j$ represent the ionic displacements, and $\omega_0$ is the phonon frequency. 
The phonons' coupling to the nematic degrees of freedom is dictated by symmetry as
\begin{eqnarray}
   H_{\rm nem} &=&
      \lambda^2 \tau_z (u_x^{2} - u_y^{2}) -\sigma_{\rm ext} \tau_z +\frac{\Omega}{2}\tau_x .
\label{eq: Hnem}
\end{eqnarray}
$\lambda^2$ is the strength of the symmetry-allowed coupling between Ising-nematic degree (transforming according to {\bg}) to the doubly degenerate \eg phonons. The Pauli matrices $\tau_{\alpha}$ act in the space of the nematic pseudospin. $\sigma_{\rm ext} $ is proportional to externally applied stress that explicitly breaks the four-fold symmetry. It acts like an conjugate field of the pseudospins. Finally
$\Omega$ is the rate at which the nematic state undergoes quantum fluctuations. Without going into the details of a microscopic description, the rate can be thought of as a renormalized tunneling rate $\Omega_0 \rightarrow \Omega = \Omega_0e^{-\int_0^{\omega_c}\frac{{\rm Im} \Gamma(\omega)}{(\omega+\Omega_0)^2}d\omega}$ rooted in a more complex dynamic nematic susceptibility of the form $\chi_{\rm nem} (\Omega)= (\Omega_0^2-\omega^2+\Gamma(\omega))^{-1}$, similar to other pseudospin problems~\cite{Leggett_RMP1987}.
While parameters $\lambda$ and $\Omega$ enter as potentially $T$-dependent phenomenological parameters, $H_{\rm nem}+H_{E_g}$ is a non-trivial many-body problem that describes electronic nematic modes and \eg phonons that undergo nonlinear interactions. It is however possible to obtain an exact solution and to determine all many-body eigenstates and hence the \eg phonon spectrum renormalized by dynamic nematic fluctuations. Our exact solution also reveals that while the impact of the nematic mode on the \eg phonons is very strong at strong coupling, the \bg nematic spectrum is only weakly modified. Hence, we focus on the former.
 
\begin{figure*}[!htbp]
 \centering
\includegraphics[width=\columnwidth]{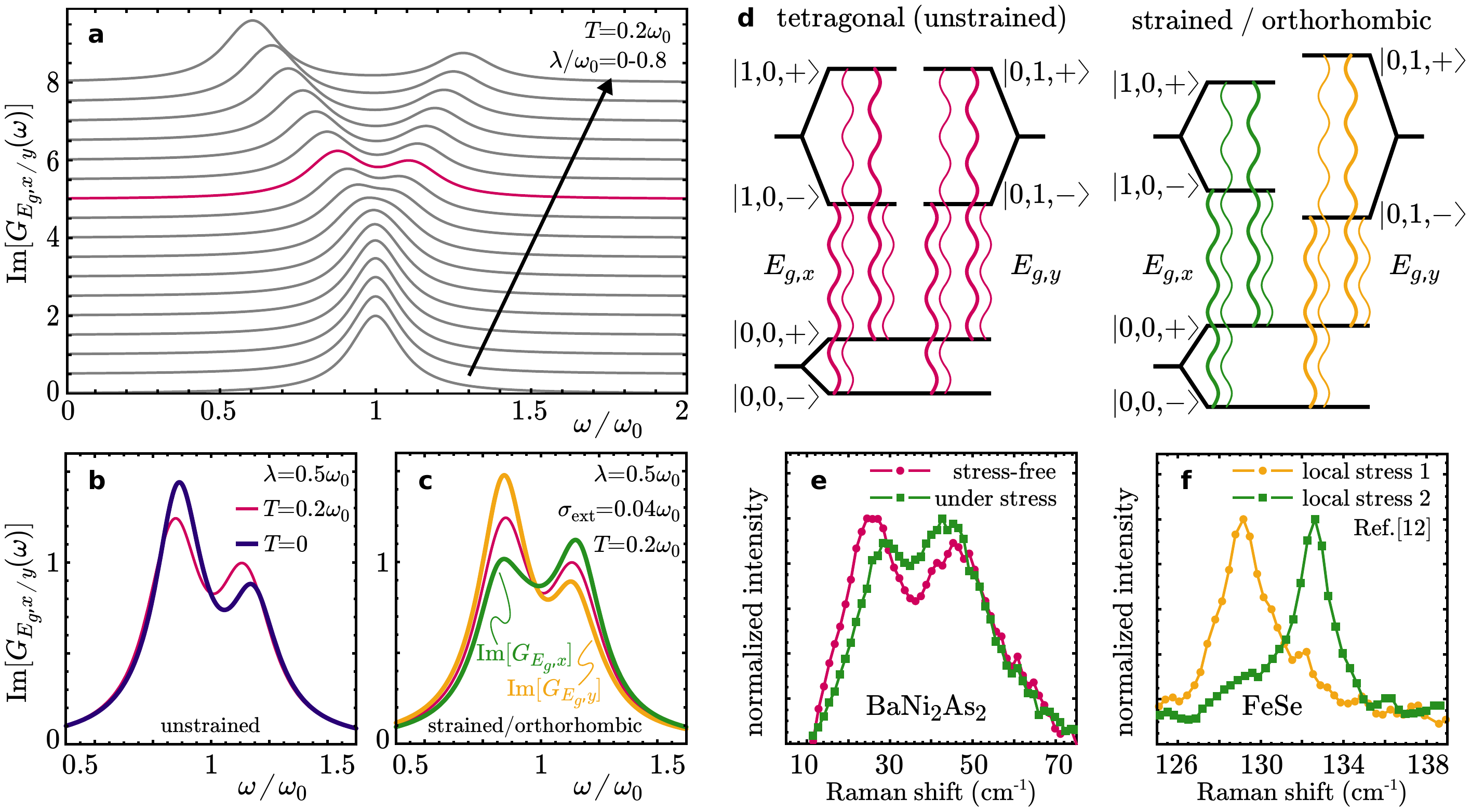}
\caption{\label{fig4} {
\textbf{Theory}}
(\textbf{a}) Calculation of the \eg Raman response using a simple model of two degenerate harmonic oscillators coupled via $\lambda$ to a fluctuating nematic degree of freedom. Details of the calculation are laid out in the Supplementary Information. Parameters for the $B_{1g}$ fluctuation frequency and temperature are chosen $\Omega = \omega_0/20$, $T = \omega_0/5$. (\textbf{b}) The weight distribution of the peak splitting depends on the relative energy scales in the problem, as illustrated for two different temperatures $T \ll \Omega$ and $\Omega \ll T \ll \omega_0$. (\textbf{c}) In the disordered case with equal peak splitting the degeneracy of the two Raman responses can be lifted by applying a conjugate external strain $\sigma_{\rm ext}$. (\textbf{d}) Schematic of the allowed transitions that cause the peak-splitting of the Raman signal even in the tetragonal state. (\textbf{e}) Raman response of \bna stress-free and under uniaxial stress and comparison to local stress dependence (\textbf{f}) of FeSe [Data from Ref.~\onlinecite{Hu_PRB2016}, plotted with permission from the authors].
}
\end{figure*}

Before discussing the results for the phonon spectrum in the tetragonal phase, relevant for \bna, we briefly discuss what happens in a $C_4$-symmetry broken state where the nematic Ising order parameter $\langle\tau_z\rangle$ is finite. There, the degeneracy of the \eg modes is lifted with their energy  becoming $\sqrt{\Omega^2 \pm \langle \tau_z \rangle \lambda^2}$. In the limit where $\lambda^2 \langle \tau_z \rangle \ll \Omega^2$, the splitting is directly given by $\frac{\lambda^2}{\Omega^2} \langle \tau_z \rangle$ and therefore provides a direct measurement of the coupling strength between the lattice and the nematic degrees of freedom. This corresponds to the behavior observed in the Fe-based superconductors\cite{Hu_PRB2016}. 
More surprising is our finding for the disordered phase in which $\langle\tau_z\rangle = 0$, shown in Fig.~\ref{fig4}-a for $\Omega \ll T \ll \omega_0$. With increasing coupling strength, we find a redistribution of the spectral weight in the phonon spectra, yielding a pronounced splitting in the strong coupling regime ($\lambda \geq 0.4 \omega_{0}$) even though the degeneracy of the \eg modes has not been lifted. The details of the spectral weight of the two peaks are affected by the ratio of the relevant energy scales, as shown in  Fig.~\ref{fig4}-b. In particular, at low temperature $T \ll \Omega$ the reduction of phase space produces an imbalance in the total weight of the two peaks.

The main message of the analysis presented here is that, even without nematic symmetry breaking, the fluctuations between the $|\!\Uparrow\rangle$ and $|\!\Downarrow \rangle$ states can yield a dynamic splitting of each \eg mode, which in turn can be seen as a direct signature of strong nemato-elastic coupling. A key prediction of the model is that in the dynamical regime, unlike in the ordered state, both $u_x$ and $u_y$ modes exhibit the same split spectrum. 
If we apply external stress $\sigma_{\rm ext}$  the degeneracy of the two $E_g$ phonons is lifted. We then observe merely a gradual transfer of weight between the split peaks, see Fig.~\ref{fig4}-c.
The relevant processes for the Raman response are given in Fig.~\ref{fig4}-c, both for the degenerate disordered state and for the degeneracy-lifted orthorhombic state. For a more detailed account of the model used here we refer to the Supplementary Information.

To experimentally investigate the symmetry-lifting for \bna, we performed measurements under strain, in a comparative study with FeSe. Quite generally, Fe-based superconductors are particularly soft and can be detwinned with very modest stress. This can be seen in a Raman experiment through the suppression in the intensity of one of the two degenerate B$_{2g}$ or B$_{3g}$  modes~\cite{Baum_PRB2018, Zhang_PRB2016, Ren_PRL2016, Hu_PRB2016}. Here, we used the approach proposed in Ref.~\cite{He_NatCom2017}, gluing a \bna sample onto a glass-fiber reinforced plastic substrate with the edges of the tetragonal unit cell aligned with the fibers. The resulting symmetry breaking strain is estimated to 0.4\% at 150K, which decreases the triclinic transition temperature by about 5K and yields a small but measurable shift of the \ag phonon of $\sim$ 0.5\icm (Supplemmentary Information). In sharp contrast to the Fe-based compounds (Fig.~\ref{fig4}-f), the ratio between the two \eg features barely changes in \bna (Fig.~\ref{fig4}-e), except for the expected transfer of spectral weight, hereby confirming the dynamical nature of the nematic electronic phase of \bna.
 
\begin{figure*}[!htbp]
 \centering
\includegraphics[width=.8\textwidth]{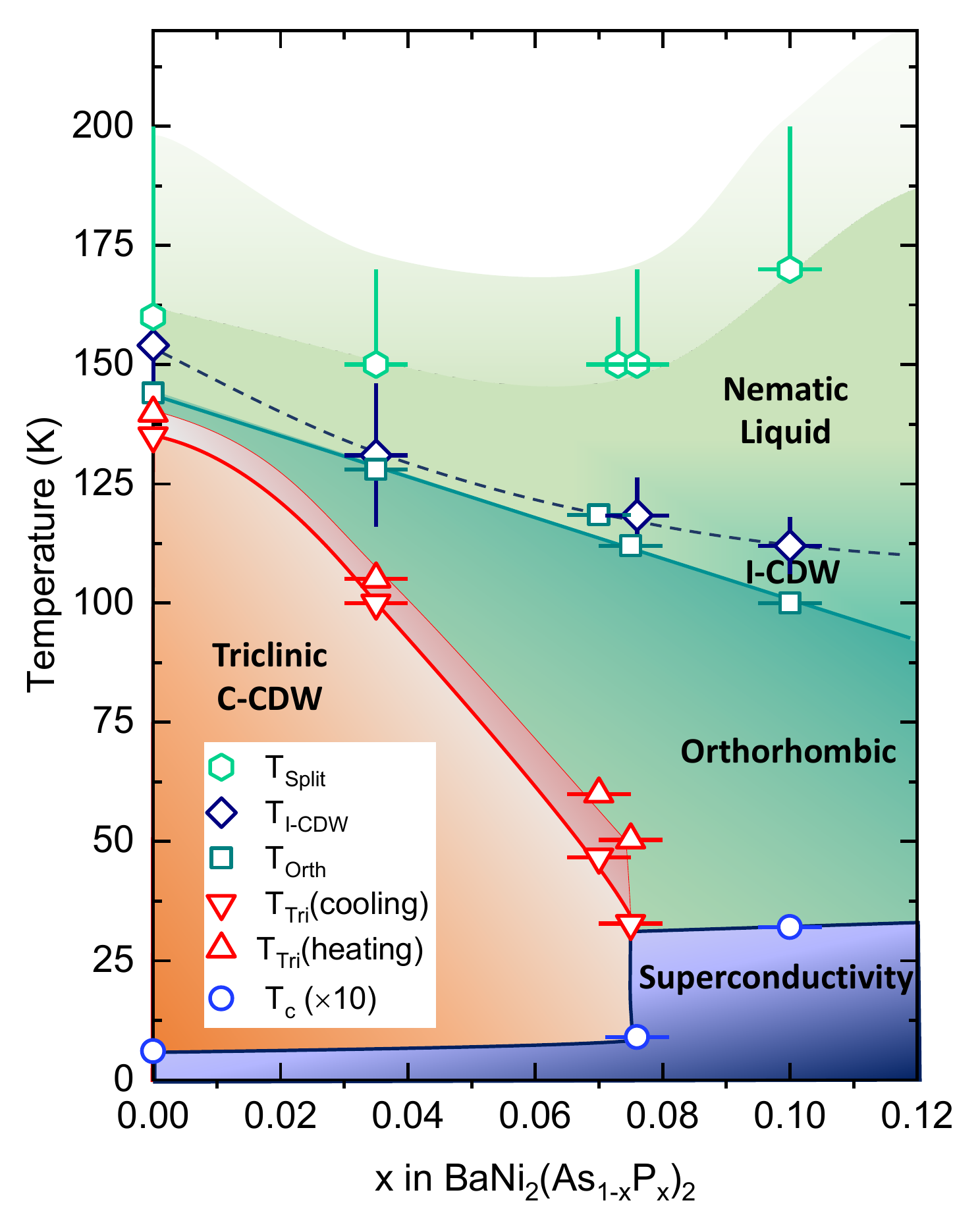}
\caption{\label{fig5} {\textbf{Phase diagram}} (Proposed phased diagram for \bnap. The transition temperatures for the triclinic and orthorhombic phases are determined from transport and thermal expansion measurements. The superconducting transition temperature is measured by specific heat (Supplementary Information). The onset of the C-CDW seen with XRD coincides with the triclinic transition, whereas the I-CDW appears above the orthorhombic distortion. The splitting of the \ega Raman phonon is indicated by $T_{Split}$, to which we have added a vertical error bar corresponding to the onset of the broadening of the mode (Supplementary Information).}
\label{fig:phasediag}
\end{figure*}

\paragraph*{Discussion} 
We summarize our findings on the phase diagram shown in Fig.~\ref{fig5} in which we report the doping dependence of the triclinic transition temperature and of the orthorhombic transition as determined from a combination of XRD, Raman, resistivity, specific heat and dilatometry experiments. The main result of this study is the pronounced broadening and splitting of the \eg modes which occurs at temperatures significantly larger than that of static structural distortions and/or of the apparition of CDW orders.
This can be explained by a strong symmetry-allowed coupling between the lattice and dynamic  $B_{1g}$ nematic fluctuations. The \eg phonons follow the dynamics of the Ising variable, likely related to resonant transitions between distinct electronic orbital states, which causes a splitting in the phonon spectrum. For this splitting no symmetry breaking is necessary. Just like the splitting of the electronic spectrum in a Mott-Hubbard insulator into the two Hubbard bands can occur without symmetry breaking, here we capture the nematic system in terms of a purely dynamical description. The analogy to spin-liquid states in Mott insulators without magnetic order may carry over to nematic liquids and give rise to related phenomena, such as unconventional superconductivity or exotic quantum states with long-range entanglement. It strongly supports the view that the much-enhanced superconducting transition temperature of these materials upon doping is closely tied to the emergence of dynamic nematic fluctuations uncovered in our measurements.

\section*{Methods}

\paragraph*{Single Crystal Growth.}
Single crystals of \bnap~ were grown using a self-flux method. 
NiAs binary was synthesized by mixing the pure elements Ni (powder, Alfa Aesar $99.999\%$) and As (lumps, Alfa Aesar $99.9999\%$) 
that were ground and sealed in a fused silica tube and annealed for 20 hours at $730\, ^\circ{\rm C}$. 
All sample handlings were performed in an argon glove box (O$_2$ content $< 0.7 \, {\rm ppm}$). 
For the growth of \bnap, a ratio of Ba:NiAs:Ni:P $= 1:4(1-x):4x:4x$ was placed in an alumina tube, which was sealed in an evacuated quartz ampule (\textit{i.e.} 10$^{-5}$ mbar). 
The mixtures were heated to $500\, ^\circ{\rm C}-700\, ^\circ{\rm C}$ for 10 h, followed by heating slowly to a temperature of $1100\, ^\circ{\rm C}-1150\, ^\circ{\rm C}$, soaked for $5\,{\rm h}$, and subsequently cooled to $995\, ^\circ{\rm C}-950\, ^\circ{\rm C}$ 
at the rate of $0.5\, ^\circ {\rm C}/{\rm h}$ to $1\, ^\circ {\rm C}/{\rm h}$, depending on the phosphorus content used for the growth. At $950\, ^\circ{\rm C}-995\, ^\circ{\rm C}$, the furnace was canted to remove the excess flux, followed by furnace cooling. 
Plate-like single crystals with typical sizes $3 \times 2 \times 0.5\, {\rm mm}^3$ were easily removed from the remaining ingot. The crystals were brittle having shiny brass-yellow metallic lustre. Electron micro-probe analysis of the \bnap~crystals was performed using a compact scanning electron microscope (SEM) (see Supplementary Information). The  energy dispersive x-ray (EDX) analysis on the \bnap~crystals revealed phosphorus content $x = 0.035 \pm 0.005$, $0.076\pm 0.005$, and $0.10\pm 0.005$.

\paragraph*{Single crystal X-ray diffraction.}
The phosphorus concentrations of the investigated samples were further confirmed by structural refinement from x-ray diffraction using an image plate system 
as described in \cite{Merz_arxiv2020}.
Detailed temperature dependencies of the I- and C-CDW superstructure peaks were obtained using a four circle diffractometer.
The samples were cooled under vacuum in a DE-202SG/700K closed-cycle cryostat from ARS, surrounded by a Beryllium dome. 
The incoming beam was generated from a Molybdenum X-ray tube with a voltage of 50 kV and a current of 40 mA. The beam was collimated
and cleaned up by a 0.8 mm pinhole before hitting the samples. We specifically followed the superstructure reflections close to the (4, 1, 1) and 
(1, 0, 3) Bragg peaks for the I- and C-CDW, respectively.

\paragraph*{Polarization-resolved confocal Raman scattering.}
Confocal Raman scattering experiments were performed with a Jobin-Yvon LabRAM HR Evolution spectrometer in backscattering geometry, with a laser power of ${\leq}0.8$\,mW that was focused on the sample with a $50${$\times$} magnification long-working-distance ($10.6$\,mm) objective. The laser spot size was $\approx2$\,$\rm{{\mu}m}$ in diameter. Low-resolution mode (1.54 \icm) of the spectrometer with 600 grooves/mm was used to maximize the signal output. For the phonon measurements, a He-Ne laser ($\lambda=632.8$\,nm) was used as the incident source, whereas the electronic background was best observed using the 532 nm line of a Nd:YAG solid state laser.
Direct comparison of the structural phase transition temperature upon cooling as measured in specific heat and Raman indicated a laser-induced heating limited to less than 2K.
The Raman spectra were Bose corrected and the phonons analyzed using a damped harmonic oscillator profile (with the exception of the $A_{1g}$ mode that displayed a Fano asymmetry and was treated accordingly).

Additional details on the experiment, phonon calculations or on the theoretical model presented here are given in the Supplementary Information.

\noindent\textbf{Acknowledgements}
We are grateful to P. Abbamonte and J. P. Paglione for valuable discussions. 
The contribution from M.M. was supported by the Karlsruhe Nano Micro Facility for Information (KNMFi). We further acknowledge the support of the KNMFi and Dr. Torsten Scherer for the EDX measurements.
We acknowledge the funding by the Deutsche Forschungsgemeinschaft (DFG; German Research Foundation) Project-ID 422213477 - TRR 288 and support by the state of Baden-W\"{u}rttemberg through bwHPC.

\noindent\textbf{Author contributions}
M.L.T. conceived and supervised the project. Y.Y. and T. L. acquired and analyzed the Raman scattering data. A.A.H. and T. L. grew the single crystals. T. L. , S.M.S., F. W. and M.M. carried out XRD experiments and analysis. M. F. performed transport experiments. C. M. performed dilatometry experiments. K. W. performed specific heat experiments. R.H. performed first-principle calculations and R. W. and J. S. developed the theoretical model. 
M.L.T, R. W. and J. S. wrote the manuscript with inputs from all the co-authors. 

\noindent\textbf{Data availability}
The data that support the findings of this study are available from the corresponding author,  M.~L.~T.(\href{mailto:matthieu.letacon@kit.edu}{matthieu.letacon@kit.edu}), upon reasonable request.

\noindent\textbf{Competing interests}
The authors declare no competing interests.

\newpage

%
%

\bigskip


\newpage
\begin{thebibliography}{99}

\bibitem{Fernandes_ARCMP2019} Fernandes, R. M., Orth, P. P. \& Schmalian, J. Intertwined Vestigial Order in Quantum Materials: Nematicity and Beyond. \textit{Annual Review of Condensed Matter Physics} \textbf{10}, 133-154 (2019).

\bibitem{Lederer_PRL2015} Lederer, S., Schattner, Y., Berg, E. \& Kivelson, S. A. Enhancement of Superconductivity near a Nematic Quantum Critical Point. \textit{Phys. Rev. Lett.} \textbf{114}, 097001 (2015).

\bibitem{Boehmer_CRP2016} B\"{o}hmer, A. E. \& Meingast, C. Electronic nematic susceptibility of iron-based superconductors. \textit{Comptes Rendus Physique} \textbf{17}, 90-112 (2016).

\bibitem{Chu_PRB2009} Chu, J.-H., Analytis, J. G., Kucharczyk, C. \& Fisher, I. R. Determination of the phase diagram of the electron-doped superconductor Ba(Fe$_{1 - x}$Co$_{x}$)$_{2}$As$_{2}$. \textit{Phys. Rev. B} \textbf{79}, 014506-014506 (2009).

\bibitem{Weber_PRB2018} Weber, F. \textit{et al.} Soft phonons reveal the nematic correlation length in $\mathrm{Ba}{({\mathrm{Fe}}_{0.94}{\mathrm{Co}}_{0.06})}_{2}{\mathrm{As}}_{2}$. \textit{Phys. Rev. B} \textbf{98}, 014516 (2018).

\bibitem{Merritt_PRL2020} Merritt, A. M. \textit{et al.} Nematic Correlation Length in Iron-Based Superconductors Probed by Inelastic X-Ray Scattering. \textit{Phys. Rev. Lett.} \textbf{124}, 157001 (2020).

\bibitem{Wu_PRL2021} Wu, S. \textit{et al.} Short-Range Nematic Fluctuations in ${\mathrm{Sr}}_{1\ensuremath{-}x}{\mathrm{Na}}_{x}{\mathrm{Fe}}_{2}{\mathrm{As}}_{2}$ Superconductors. \textit{Phys. Rev. Lett.} \textbf{126}, 107001 (2021).

\bibitem{Chauviere_PRB2009} Chauvi\`{e}re, L. \textit{et al.} Doping dependence of the lattice dynamics in  Ba(Fe$_{1 - x}$Co$_{x}$)$_{2}$As$_{2}$ studied by Raman spectroscopy. \textit{Phys. Rev. B} \textbf{80}, 094504-094506 (2009).

\bibitem{Ren_PRL2016} Ren, X. \textit{et al.} Nematic Crossover in ${\mathrm{BaFe}}_{2}{\mathrm{As}}_{2}$ under Uniaxial Stress. \textit{Phys. Rev. Lett.} \textbf{115}, 197002 (2015).

\bibitem{Baum_PRB2018} Baum, A. \textit{et al.} Interplay of lattice, electronic, and spin degrees of freedom in detwinned ${\mathrm{BaFe}}_{2}{\mathrm{As}}_{2}$: A Raman scattering study. \textit{Phys. Rev. B} \textbf{98}, 075113 (2018).

\bibitem{Zhang_PRB2016} Zhang, W. L. \textit{et al.} Stress-induced nematicity in ${\mathrm{EuFe}}_{2}{\mathrm{As}}_{2}$ studied by Raman spectroscopy. \textit{Phys. Rev. B} \textbf{94}, 014513 (2016).

\bibitem{Hu_PRB2016} Hu, Y. \textit{et al.} Nematic magnetoelastic effect contrasted between $\mathrm{Ba}({\mathrm{Fe}}_{1\ensuremath{-}x}{\mathrm{Co}}_{x}{)}_{2}{\mathrm{As}}_{2}$ and FeSe. \textit{Phys. Rev. B} \textbf{93}, 060504 (2016).

\bibitem{Kudo_PRL2012} Kudo, K. \textit{et al.} Giant Phonon Softening and Enhancement of Superconductivity by Phosphorus Doping of ${\mathrm{BaNi}}_{2}{\mathrm{As}}_{2}$. \textit{Phys. Rev. Lett.} \textbf{109}, 097002 (2012).

\bibitem{Ronning_JPCM2008} Ronning, F. \textit{et al.} The first order phase transition and superconductivity in BaNi$_2$As$_2$ single crystals. \textit{Journal of Physics: Condensed Matter} \textbf{20}, 342203 (2008).

\bibitem{Subedi_PRB2008} Subedi, A. \& Singh, D. J. Density functional study of ${\text{BaNi}}_{2}{\text{As}}_{2}$: Electronic structure, phonons, and electron-phonon superconductivity. \textit{Phys. Rev. B} \textbf{78}, 132511 (2008).

\bibitem{Kurita_PRL2009} Kurita, N. \textit{et al.} Low-Temperature Magnetothermal Transport Investigation of a Ni-Based Superconductor ${\mathrm{BaNi}}_{2}{\mathrm{As}}_{2}$: Evidence for Fully Gapped Superconductivity. \textit{Phys. Rev. Lett.} \textbf{102}, 147004 (2009).

\bibitem{Eckberg_NatPhys2020} Eckberg, C. \textit{et al.} Sixfold enhancement of superconductivity in a tunable electronic nematic system. \textit{Nat. Phys.} \textbf{16}, 346-350 (2020).

\bibitem{Frachet2022} Frachet, M. \textit{et al.}, Elastoresistivity in the incommensurate charge density wave phase of BaNi$_2$(As$_{1-x}$P$_x$)$_2$. Preprint at \href{https://arxiv.org/abs/2207.02462 (2022)}{https://arxiv.org/abs/2207.02462 (2022)}.

\bibitem{Eckberg_PRB2018} Eckberg, C. \textit{et al.} Evolution of structure and superconductivity in ${\mathrm{Ba}(\mathrm{Ni}}_{1\ensuremath{-}x}{\mathrm{Co}}_{x}{)}_{2}{\mathrm{As}}_{2}$. \textit{Phys. Rev. B} \textbf{97}, 224505 (2018).

\bibitem{Lee_PRL2019} Lee, S. \textit{et al.} Unconventional Charge Density Wave Order in the Pnictide Superconductor $\mathrm{Ba}({\mathrm{Ni}}_{1\ensuremath{-}x}{\mathrm{Co}}_{x}{)}_{2}{\mathrm{As}}_{2}$. \textit{Phys. Rev. Lett.} \textbf{122}, 147601 (2019).

\bibitem{Merz_arxiv2020}  Merz, M. \textit{et al.} Rotational symmetry breaking at the incommensurate charge-density-wave transition in $\mathrm{Ba}{(\mathrm{Ni},\mathrm{Co})}_{2}{(\mathrm{As},\mathrm{P})}_{2}$: Possible nematic phase induced by charge/orbital fluctuations. \textit{Phys. Rev. B} \textbf{104}, 184509 (2021).

\bibitem{Meingast2022} Meingast, C. \textit{et al.}, Charge-density-wave transitions, phase diagram, soft phonon and possible electronic nematicity: a thermodynamic investigation of BaNi$_2$(As,P)$_2$. Preprint at \href{https://arxiv.org/abs/2207.02294  (2022)}{https://arxiv.org/abs/2207.02294 (2022)}.

\bibitem{Grigorev_arxiv2021} Pokharel, A. R. \textit{et al.} Dynamics of collective modes in an unconventional charge density wave system BaNi$_{2}$As$_{2}$. \textit{Communications Physics} \textbf{5}, 141 (2022).

\bibitem{Lee_arxiv2021} Lee, S. \textit{et al.} Multiple charge density waves and superconductivity nucleation at antiphase domain walls in the nematic pnictide 
 Ba$_{1-x}$Sr$_{x}$Ni$_{2}$As$_{2}$. \textit{Phys. Rev. Lett.} \textbf{127}, 027602 (2021).

\bibitem{Lederer_PRR2020} Lederer, S., Berg, E. \& Kim, E.-A. Tests of nematic-mediated superconductivity applied to ${\mathrm{Ba}}_{1\ensuremath{-}x}{\mathrm{Sr}}_{x}{\mathrm{Ni}}_{2}{\mathrm{As}}_{2}$. \textit{Physical Review Research} \textbf{2}, 023122 (2020).

\bibitem{Devereaux_RMP2007} Devereaux, T. P. \& Hackl, R. Inelastic light scattering from correlated electrons. \textit{Reviews of Modern Physics} \textbf{79}, 175-159 (2007).

\bibitem{Le Tacon_NaturePhysics06} Le Tacon, M. \textit{et al.} Two energy scales and two distinct quasiparticle dynamics in the superconducting state of underdoped cuprates. \textit{Nature Physics} \textbf{2}, 537 (2006).

\bibitem{Gallais_PRL2013} Gallais, Y. \textit{et al.} Observation of Incipient Charge Nematicity in ${\mathrm{Ba}(\mathrm{Fe}{}_{1\ensuremath{-}X}\mathrm{Co}{}_{X})}_{2}\mathrm{As}{}_{2}$. \textit{Phys. Rev. Lett.} \textbf{111}, 267001 (2013).

\bibitem{Fano_PR1961} Fano, U. Effects of Configuration Interaction on Intensities and Phase Shifts. \textit{Physical Review} \textbf{124}, 1866-1878 (1961).

\bibitem{Gallais_CR2016} Gallais, Y. \& Paul, I. Charge nematicity and electronic Raman scattering in iron-based superconductors. \textit{Comptes Rendus Physique} \textbf{17}, 113-139 (2016).

\bibitem{Sen_NatCom2020} Sen, K. \textit{et al.} Strange semimetal dynamics in SrIrO$_3$. \textit{Nat. Comm.} \textbf{11}, 4270 (2020).

\bibitem{Kretzschmar_NatPhys2016} Kretzschmar, F. \textit{et al.} Critical spin fluctuations and the origin of nematic order in Ba(Fe$_{1-x}$Co$_x$)$_2$As$_2$. \textit{Nat. Phys.} \textbf{12}, 560-563 (2016).

\bibitem{Silbey_JCP1983} Silbey, R. \& Harris, R. A., Variational calculation of the dynamics of a two level system interacting with a bath, \textit{J. Chem. Phys.} {\bf 80}, 2615 (1983).

\bibitem{Silbey_JCP1989}  Tunneling of molecules in low-temperature media: an elementary description, \textit{J. Chem. Phys.} {\bf 93}, 7062 (1989).

\bibitem{Leggett_RMP1987} Leggett,  A. J., Chakravarty, S. Dorsey, A. T., Fisher, M. P. A., Garg, A.,  and  Zwerger, W., Dynamics of the dissipative two-state system, \textit{Rev. Mod. Phys.}  \textbf{59}, 1 (1987).

\bibitem{Costi_PRL1996} Costi, T. A. \& Kieffer, C., Equilibrium Dynamics of the Dissipative Two-State System, \textit{Phys. Rev. Lett.} {\bf 76}, 1683 (1996).

\bibitem{He_NatCom2017} He, M. \textit{et al.} Dichotomy between in-plane magnetic susceptibility and resistivity anisotropies in extremely strained BaFe$_2$As$_2$. \textit{Nat. Comm.} \textbf{8}, 504 (2017).

\bibitem{SOC} Heid, R. Bohnen, K.-P., Sklyadneva, I. Yu. Chulkov, E. V. \textit{Physical Review B} \textbf{81}, 174527 (2010) 

\end{thebibliography}
\end{document}